\documentclass[11pt,a4paper]{article}
\pdfoutput=1

\usepackage{jheppub}
\usepackage{latexsym}
\usepackage{revsymb}
\usepackage{multirow}

\usepackage{graphicx}
\usepackage{epsfig}  
\usepackage{epsf}    
\usepackage{dcolumn}
\usepackage{bm}
\usepackage{dcolumn}
\usepackage{textcomp}
\usepackage{float}
\usepackage{subfig}
\usepackage[]{hyperref}
  
\hypersetup{
  bookmarks=true,         
  unicode=false,          
  pdftoolbar=true,        
 pdfmenubar=true,        
 pdffitwindow=true,     
 pdfstartview={FitH},    
 pdfsubject={Neutrino Oscillations Phenomenology},   
 pdfnewwindow=true,      
 pdfcreator={RevTeX},
 colorlinks=true,       
 linkcolor=red,          
 citecolor=blue,        
filecolor=black,      
 urlcolor=blue,           
  }

\usepackage{hypcap}

\newcommand{\beq}{\begin{equation}}
\newcommand{\eeq}{\end{equation}}
\newcommand{\beqa}{\begin{eqnarray}}
\newcommand{\eeqa}{\end{eqnarray}}

\newcommand{\ty}{{\theta_{13}}}
\newcommand{\tz}{{\theta_{23}}}

\newcommand{\dcp}{\delta_{\mathrm{CP}}}
\newcommand{\nova}{NO$\nu$A~}
\newcommand{\pmue}{P(\nu_\mu \rightarrow \nu_e)}
\newcommand{\dcpt}{\delta_{CP}{\mbox {(true)}}}

\def\nue{{\nu_e}}
\def\anue{{\bar\nu_e}}

\preprint{IFIC/12-60, EURONU-WP6-12-52}

\title{Potential of optimized NO$\nu$A for large $\theta_{13}$ \& combined performance with a LArTPC \& T2K}

\author[a]{Sanjib Kumar Agarwalla,}
\author[b]{Suprabh Prakash,} 
\author[b,c]{Sushant K. Raut,} 
{\author[b,d]{S. Uma Sankar$\,$}

\affiliation[a]{Instituto de F\'{\i}sica Corpuscular, CSIC-Universitat de Val\`encia, \\
       Apartado de Correos 22085, E-46071 Valencia, Spain}
\affiliation[b]{Department of Physics, Indian Institute of Technology Bombay, Mumbai 400076, India}
\affiliation[c]{Physical Research Laboratory, Ahmedabad 380009, India}
\affiliation[d]{Department of Theoretical Physics, Tata Institute of Fundamental Research, \\
      Mumbai 400005, India}

\emailAdd{Sanjib.Agarwalla@ific.uv.es}
\emailAdd{suprabh@phy.iitb.ac.in}
\emailAdd{sushant@phy.iitb.ac.in}
\emailAdd{uma@phy.iitb.ac.in}

\abstract
{
NO$\nu$A experiment has reoptimized its event selection criteria in 
light of the recently measured moderately large value 
of $\theta_{13}$. We study the improvement in the sensitivity to the 
neutrino mass hierarchy and to leptonic CP violation due to 
these new features. For favourable values of $\dcp$, \nova sensitivity to mass hierarchy and 
leptonic CP violation is increased by 20\%. Addition of 5 years of neutrino data 
from T2K to \nova more than doubles the range of $\dcp$
for which the leptonic CP violation can be discovered, compared 
to stand alone NO$\nu$A. But for unfavourable values of $\dcp$, 
the combination of NO$\nu$A and T2K are not 
enough to provide even a 90\% C.L. hint of hierarchy discovery. 
Therefore, we further explore the improvement in the hierarchy
and CP violation sensitivities due to the addition of a $10\,
\mathrm{kt}$ liquid argon detector placed 
close to NO$\nu$A site. The capabilities of such a detector 
are equivalent to those of NO$\nu$A in all respects. We find that 
combined data from $10\,\mathrm{kt}$ liquid argon detector 
(3 years of $\nu$ + 3 years of $\bar\nu$ run), 
NO$\nu$A (6 years of $\nu$ + 6 years of $\bar\nu$ run) and T2K 
(5 years of $\nu$ run) can give a close to $2\sigma$ hint of 
hierarchy discovery for all values of $\dcp$. With this combined 
data, we can achieve CP violation discovery at 95\% C.L. 
for roughly 60\% values of $\dcp$.        
}


\begin{document}
\maketitle
\flushbottom

\section{Introduction}
\label{sec:intro}

The recent measurements of $\theta_{13}$ by the reactor neutrino experiments Double Chooz~\cite{Abe:2011fz,Abe:2012tg}, 
Daya Bay~\cite{An:2012eh,Kyoto2012DayaBay}, and RENO~\cite{Ahn:2012nd,Kyoto2012RENO} are a very welcome news 
for future neutrino oscillation experiments. In addition to confirming the non-zero $\ty$ as hinted earlier by the accelerator 
experiments T2K~\cite{Abe:2011sj,Kyoto2012T2K} and MINOS~\cite{Adamson:2011qu,Kyoto2012MINOS}, 
these experiments have determined the value of $\ty$ quite precisely. Global fits to the world neutrino 
data~\cite{fogli2012,tortola2012} give a best fit value $\sin^2 2 \ty = 0.095$ with a $1\sigma$ uncertainty
of about $10\%$~\cite{tortola2012}. Daya Bay experiment  
is expected to reduce the uncertainty to $5\%$ level by the time it finishes running in 2016 \cite{dayabay_NF12}. 
Since the value of $\sin^2 2 \theta_{13}$ is moderately
large (in fact, it is just below the CHOOZ upper limit \cite{chooz1,chooz2}),
the current (T2K) \cite{t2k} and upcoming (NO$\nu$A) \cite{nova} 
experiments now have a chance of determining the
remaining unknowns of neutrino oscillations: {\it i.e.} (a) neutrino
mass hierarchy (equivalently the sign of $\Delta m^2_{31}=m^2_3 - m^2_1$) 
and (b) existence of CP violation in the leptonic sector.

CP violation in the leptonic sector has drawn tremendous interest because of the possibility of leptogenesis 
leading to baryogenesis and the baryon asymmetry of the universe~\cite{Fukugita:1986hr}. 
Leptogenesis requires the existence of CP violation in the leptonic sector; see~\cite{DiBari:2012fz} for a recent review.
The possible connection between leptogenesis and neutrino oscillations has been discussed 
in~\cite{Joshipura:2001ui,Endoh:2002wm,Pati:2003gv}. It is likely that the CP violating phase in neutrino oscillations 
is not directly related to the CP violation leading to leptogenesis. But a demonstration of CP violation in neutrino 
oscillations provides a crucial guidepost for models of leptonic CP violation and leptogenesis.

The experiments \nova and T2K expect to achieve the above goals 
by measuring the $\nu_\mu \rightarrow \nu_e$ oscillation probability
$P_{\mu e}$ and its charge conjugate $\bar{\nu}_\mu \rightarrow
\bar{\nu}_e$ oscillation probability $P_{\bar{\mu} \bar{e}}$. 
Since these experiments have moderately long baselines (295 km
for T2K and 810 km for NO$\nu$A), the matter effects due to neutrino
propagation through Earth \cite{msw1} are important, 
especially for \nova \cite{PRD61.013003}. The matter term 
modifies $P_{\mu e}$ (and also $P_{\bar{\mu} \bar{e}}$) differently for normal hierarchy (NH 
where $\Delta m^2_{31}$ is positive) and for inverted hierarchy
(IH where $\Delta m^2_{31}$ is negative). Thus these experiments
are capable of distinguishing between the two hierarchies.
The matter term also 
changes sign when we switch from neutrino mode to anti-neutrino
mode. Hence the matter effects induce a CP like change in the 
oscillation probabilities. Thus we have an entanglement of
changes caused by matter effects and the genuine CP violating 
phase $\dcp$. Due to this, we get two degenerate solutions 
for a given experiment: one with the right hierarchy and the
right value of $\dcp$ and one with the wrong hierarchy and 
a wrong value of $\dcp$. These degenerate solutions
can be unravelled if data from two different experiments with
different baselines is available
\cite{degeneracy1, degeneracy2, degeneracy3, degeneracy4, twobase6, twobase7}.

It was shown that the data from T2K and \nova should be synergistically
combined to obtain the best possible sensitivity to hierarchy 
\cite{menaparke}. The presently planned runs of T2K and \nova will
not be able to determine the hierarchy for the whole range of $\dcp$
\cite{hubercpv,nova_tdr,ryan_NF12}. 
Matter effects increase $\pmue$ for NH and decrease it for IH and vice verse
for $P_{\bar{\mu} \bar{e}}$. For $\dcp$ in the lower half-plane (LHP, $-180^\circ \leq \dcp \leq 0$),
$\pmue$ is larger and for $\dcp$ in the upper half-plane 
(UHP, $0 \leq \dcp \leq 180^\circ$), $\pmue$ is smaller.
Hence, for the combination (NH, LHP), the values of $\pmue$ are much 
higher than those for IH (and $P_{\bar{\mu} \bar{e}}$ values are much 
lower). Similarly, for the combination (IH, UHP), the values of $\pmue$ are
much lower than those of NH (and $P_{\bar{\mu} \bar{e}}$ values are much 
higher). Thus, LHP is the favourable half-plane for NH
and UHP is for IH \cite{novat2k}. Therefore, \nova by itself can determine the hierarchy
if $\dcp$ happens to be in the favourable half-plane \cite{nova_tdr,novat2k}. 
The range of $\dcp$ values, for which the
hierarchy can be determined, increases to some extent if the
T2K data is included. It was shown that the hierarchy can
be determined for the entire range of $\dcp$ if there is $50\%$ 
more data from \nova and twice the data from T2K \cite{novat2k}. 

In light of the moderately large value of $\theta_{13}$, \nova
experiment has reoptimized their event selection criteria, with
more events in both signal and background \cite{Kyoto2012nova,ryan}. These
new criteria improve the hierarchy determination ability significantly
in the favourable half-planes of $\dcp$ but not in the unfavourable
half-planes. The hierarchy sensitivity, for the unfavourable $\dcp$
half-planes, can be improved only with a much larger data sample. 
In this paper, we explore the expected improvement in the sensitivity to
both hierarchy and leptonic CP violation because of increased data
from experiments with 810 $\mathrm{km}$ baseline. Liquid argon time projection 
chamber (LArTPC) are emerging as a very good option for neutrino 
detector technology for future neutrino experiments both in Europe~\cite{lbno1,lbno2,icarus,incremental}
and in the USA~\cite{lbne1,duselreport,lbne3}. 
Recently, LBNE collaboration has explored the possibility of
building a 10 to 30 $\mathrm{kt}$ LArTPC at either \nova or MINOS site~\cite{lbne_NF12}. 
We calculate the combined sensitivity of a 10 $\mathrm{kt}$ LArTPC at the \nova site,
in conjunction with \nova and T2K detectors, to hierarchy and to the existence of leptonic CP violation.
 
The paper is organized as follows. In section 2, we discuss the
detectors we consider in this analysis, paying particular attention
to the improvements implemented in \nova~\cite{Kyoto2012nova} and 
contrasting them with old \nova event selection. We also detail
how these improvements have been implemented in our simulations.
We briefly describe the LArTPC and compare its properties
with those of NO$\nu$A. Sections 3 and 4 describe the event spectrum and
our numerical procedure respectively. In section 5, we present our sensitivity results
for the determination of hierarchy and the detection of CP violation for various combinations 
of experiments. Finally, in section 6, we mention our conclusions. 

\section{Detectors}

In this section, we briefly describe the main features of the detectors considered
in this report.

\subsection{NO$\nu$A}

\nova is a 14 $\mathrm{kt}$ totally active scintillator detector (TASD) at a distance 
of 810 $\mathrm{km}$ from Fermilab, at a location which is $0.8^\circ$ off-axis 
from the NuMI beam. Because of the off-axis location, the flux of the
neutrinos is reduced but is sharply peaked around 2 $\mathrm{GeV}$.
This leads to two important advantages:

\begin{itemize}

\item
The peak flux is close to the first oscillation maximum
energy of 1.7 $\mathrm{GeV}$. This leads to a large number of signal
events.

\item
The most problematic background is neutral current (NC) interactions
which mostly consists of the single $\pi^0$ production. 
However, the measured energy
of this background is shifted to values of energy below the
region where the flux is significant. Hence this background
can be rejected using a simple kinematic cut.

\end{itemize}
The experiment is scheduled to run for three years in 
neutrino mode and three years in anti-neutrino mode
with a NuMI beam power of $0.7$ MW, corresponding 
to $6\times 10^{20}$ protons on target per year.

Previously, the event selection criteria were optimized to have
the least background even under the most pessimistic 
assumption of $\sin^22\ty=0$.
This gave the best possible background rejection, at the cost of
reduced signal efficiency, which was required in the event of very
small $\sin^22\ty$. With the present, moderately large value of
$\sin^22\ty$, \nova has relaxed the cuts for the event selection
criteria which allow more signal events along with more background
events. Additional backgrounds, mostly NC, are reconstructed
at energies lower compared to the true neutrino energy and can be managed
by a kinematic cut.
The main differences between the old criteria and new 
criteria are listed below \cite{Kyoto2012nova,ryan,ryan_NF12}.

\begin{itemize}

\item
The signal efficiencies for new \nova are higher than that of old one
by roughly a factor of 2 for neutrino events. We now assume
$45\%$ signal efficiency for both neutrino and anti-neutrino events as
opposed to $26\%$ for $\nu$ and $40\%$ for $\bar{\nu}$ previously.


\item
The background acceptance has also increased. For the NC
interactions, present value is about 7 times (2\% vs. 0.3\%) for $\nu$ and 
3 times (3\% vs. 0.9\%) for $\bar{\nu}$, compared to the old criteria. 
For misidentified muons also, it is about 6 times (0.83\% vs. 0.13\%) and 
2 times (0.22\% vs. 0.13\%) the older numbers, for $\nu$ and $\bar{\nu}$ 
respectively.

\item
Earlier, the number of NC background events was moderate. In such
a situation, using a Gaussian energy resolution function to
obtain the smearing of the background events is not a bad approximation.
At present though, the NC backgrounds are higher by a factor of 5
and but their measured energy, in general, will be in a range
below the region of large flux. The NC spectrum shift to the measured energies
is implemented through migration matrices.

\item
In our analysis, we have taken care of the neutrino contamination
in the anti-neutrino beam for both appearance and disappearance
channels. It should be stressed that while anti-neutrino contamination
can be ignored in the neutrino beam, the reverse is not true.

\end{itemize}

The above optimization criteria were developed by using the
event spectra for the case of $\nu_\mu \rightarrow \nu_e$ 
vacuum oscillations with $\dcp=0$ and maximizing the signal
events while keeping the background events relatively small.
We calculated the same event spectrum under the same 
assumptions and adjusted the signal and background
efficiencies in the simulations, until we obtained the same 
number of signal events and background events as in~\cite{Kyoto2012nova}.

In Table~\ref{tab1}, we summarize the main characteristics of TASD and LArTPC
which we discuss next. 

\begin{table}[H]
{\footnotesize
     \begin{tabular}{ || l  ||p{4.7cm} || p{4.7cm}||}

         \hline
         \hline
         Detector Characteristic &LArTPC @ \nova&TASD @ \nova\\
         \hline
         \hline
         Fiducial Mass & 5 or 10 $\mathrm{kt}$ & 14 $\mathrm{kt}$ \\
         Neutrino energy threshold & 0.5 $\mathrm{GeV}$ & 0.5 $\mathrm{GeV}$ \\
         Detection efficiency
         & 85\% for $\mu^{\pm}$(CC), 80\% for $e^{\pm}$ 
         & 100\% for $\mu^{\pm}$(CCQE), 45\% for $e^{\pm}$ \\
         Energy resolution ($\mathrm{GeV}$) 
         & 0.1$\sqrt{\rm E/\mathrm{GeV}}$ for CC $\mu^{\pm}$ and $e^{\pm}$ sample
         & 0.06$\sqrt{\rm E/\mathrm{GeV}}$ for CC $\mu^{\pm}$ and 0.085$\sqrt{\rm E/\mathrm{GeV}}$ for $e^{\pm}$ sample \\
         NC backround smearing 
         & Migration Matrices 
         & Migration Matrices \\
         NC background acceptance 
         & 1\% (both $\nu$ and $\bar{\nu}$) 
         & 2\% ($\nu$), 3\% ($\bar{\nu}$) \\
         Mis-ID muons acceptance 
         & 1\% (both $\nu$ and $\bar{\nu}$) 
         & 0.83\% ($\nu$), 0.22\% ($\bar{\nu}$) \\
         Int. beam $\nu_{e}$/$\bar{\nu_{e}}$ contamination  
         & 80\% (both $\nu$ and $\bar{\nu}$) 
         & 26\% ($\nu$), 18\% ($\bar{\nu}$) \\
         Signal normalization error & 5\% & 5\%\\
         Background normalization error & 5\% & 10\% \\
         \hline
         \hline
         \end{tabular}
}
\caption{{\footnotesize Detector properties of LArTPC and TASD.}}

\label{tab1}
\end{table}


\subsection{Liquid Argon TPC}

Here we consider a 10 $\mathrm{kt}$ LArTPC constructed close to NO$\nu$A.
LArTPC has excellent particle identification and we assume a signal efficiency of $80\%$ for $e^{\pm}$ 
compared to $45\%$ for TASD. The energy resolution and background rejection for the LArTPC
and TASD are comparable. In our simulations, we have taken the efficiencies and migration matrices of 
LArTPC from~\cite{lbne3}. It is expected, of course, that such a detector will come on line much later than NO$\nu$A. 
In considering \nova + LArTPC, we assume equal 6 years $\nu$ and $\bar{\nu}$ 
runs for \nova and equal 3 years $\nu$ and $\bar{\nu}$ runs for LArTPC.
The cross-sections for LArTPC are slightly different from those of TASD. To obtain the LArTPC cross-sections, 
we have scaled the inclusive charged current (CC) cross sections by 1.06 (0.94) for the $\nu$ ($\bar{\nu}$) 
case compared to those for water~\cite{zeller,petti-zeller}.

\subsection{T2K}

T2K uses the 50 $\mathrm{kt}$ Super-Kamiokande water Cerenkov detector
(fiducial volume 22.5 $\mathrm{kt}$) as the end detector for the neutrino
beam from J-PARC. The detector is at a distance of 295 $\mathrm{km}$ from
the source at an off-axis angle of $2.5^\circ$~\cite{t2k}. The neutrino flux
is again peaked sharply at the first oscillation of $0.7$ $\mathrm{GeV}$. 
The experiment is scheduled to run for 5 years in the neutrino
mode with a power of $0.75$ MW. Because of the low energy of the
peak flux, the NC backgrounds are small and they can be rejected based
on energy cut. The signal efficiency is $87\%$. The background information 
and other details are taken from~\cite{globes_t2k4,hubercpv}.

\section{Event Rates and Spectrum}

We use GLoBES~\cite{globes1, globes2} software to simulate the data for various experiments.
For the atmospheric/accelerator neutrino parameters,  
we take the following central (true) values: 
\begin{equation}
|\Delta m^2_{\mathrm{eff}}| = 2.4 \cdot 10^{-3}\,\mathrm{eV}^2, \quad \sin^2 2 \theta_{23} = 1.0,
\label{eq:atmparam}
\end{equation}
where $\Delta m^2_{\mathrm{eff}}$ is the effective mass-squared difference
measured by the accelerator experiments in $\nu_\mu \rightarrow \nu_\mu$
disappearance channel \cite{minos}. It is related to the $\Delta m^2_{31}$
(larger) and $\Delta m^2_{21}$ (smaller) mass-square differences
through the expression~\cite{parke_defn}
\begin{equation}
\Delta m^2_{\mathrm{eff}} = \Delta m^2_{31} - 
\Delta m^2_{21} (\cos^2 \theta_{12} 
- \cos \dcp \sin \theta_{13} \sin 2 \theta_{12} \tan \theta_{23}),
\label{parkedef}
\end{equation}
where $\Delta m^2_{21}=m^2_2-m^2_1$.
The best fit for $\theta_{23}$ is taken from atmospheric neutrino
data~\cite{PRD81.092004}.
For $\theta_{13}$, we take the global best fit $\sin^2 2 \theta_{13}
=0.101$~\cite{tortola2012}\footnote{The best fit value for $\sin^2 2 \ty$ and its uncertainty
are taken from the 2nd version of~\cite{tortola2012}, dated 21 May 2012. The values quoted 
in the introduction are taken from the 3rd version of the same paper, dated 13 Aug 2012, by which 
time our calculations were finished. The 3rd version includes the data presented at the Neutrino 2012 
conference at Kyoto and hence the values have changed a little. We have checked that our conclusions
are not affected by these small changes.}. 
The uncertainties in the above parameters are taken to be
$\sigma(\sin^2 \theta_{13}) = 13\%$ \cite{tortola2012},  
$\sigma(|\Delta m^2_{\mathrm{eff}}|) = 4\%$ and  
$\sigma(\sin^2 2 \theta_{23}) = 2\%$ \cite{t2k}.
We take the solar parameters to be \cite{tortola2012}
\begin{equation} 
\Delta  m^2_{21}=7.62\cdot10^{-5}\,\mathrm{eV}^2,  \quad \sin^2 \theta_{12} =0.32.
\label{eq:solparam}
\end{equation}
We keep these parameters to be fixed throughout the 
calculation because varying them will have negligible effect. 
We also take the Earth matter density to be a constant
$2.8$ gm/cc because the variations and the uncertainties in density can be 
neglected for the baselines we consider.

Recently, MINOS experiment found the best fit for 
$\sin^2 2 \theta_{23}$ to be $0.97$ rather than $1$ \cite{Kyoto2012MINOS}.
Hence we must consider the possibility that best fit 
value for $\sin^2 \theta_{23}$ is $0.413$.
From the Daya Bay experiment, the best fit value for 
$\sin^2 2 \theta_{13}$ is $0.089$ \cite{Kyoto2012DayaBay}, which is $10\%$ 
smaller than the global best fit $0.101$ we have used.
In Appendix~\ref{appendix1}, we show how hierarchy and CP sensitivities 
change if these conservative input values are used.

\begin{figure}[H]
\centering
\includegraphics[width=0.49\textwidth]{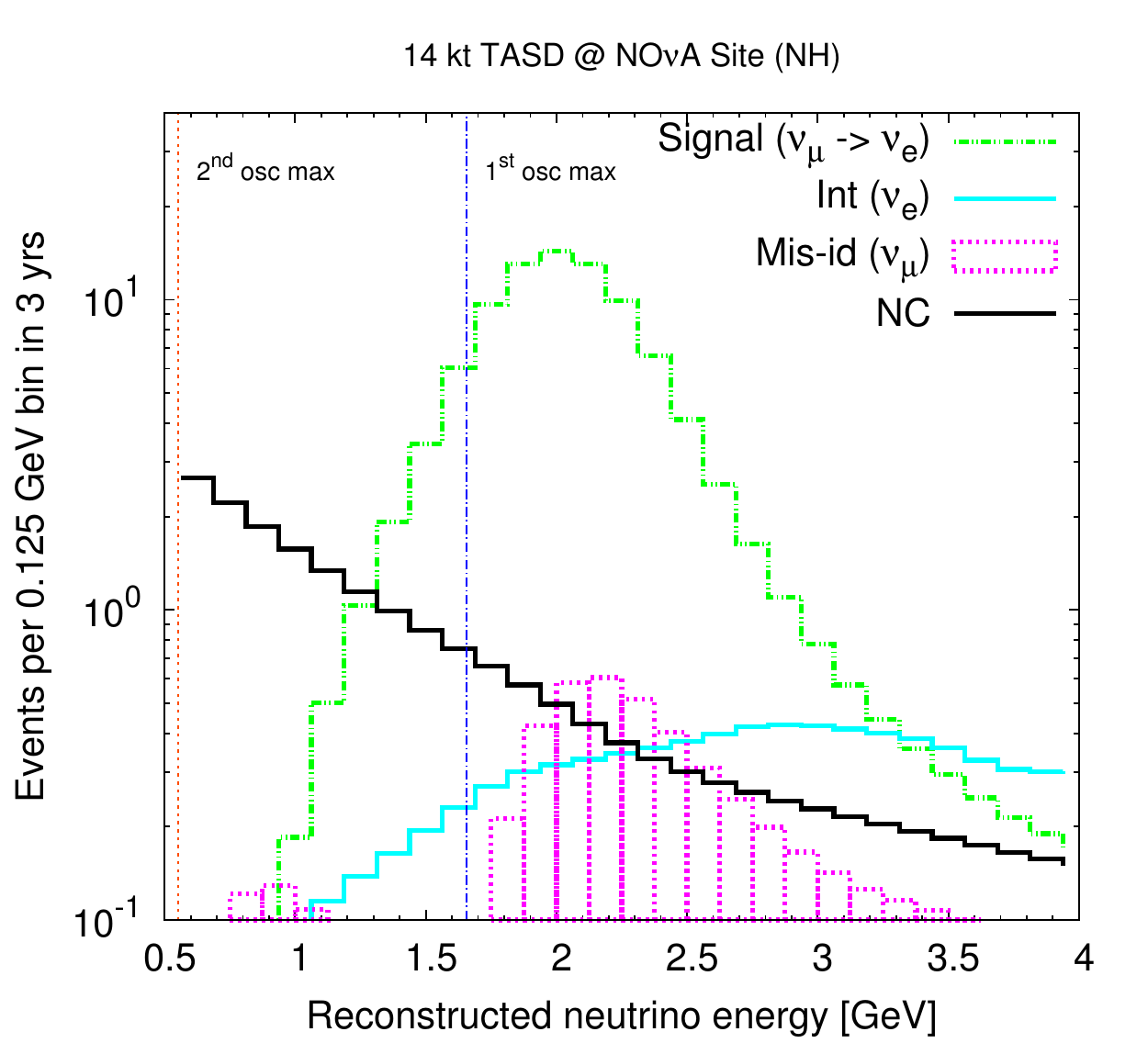}
\includegraphics[width=0.49\textwidth]{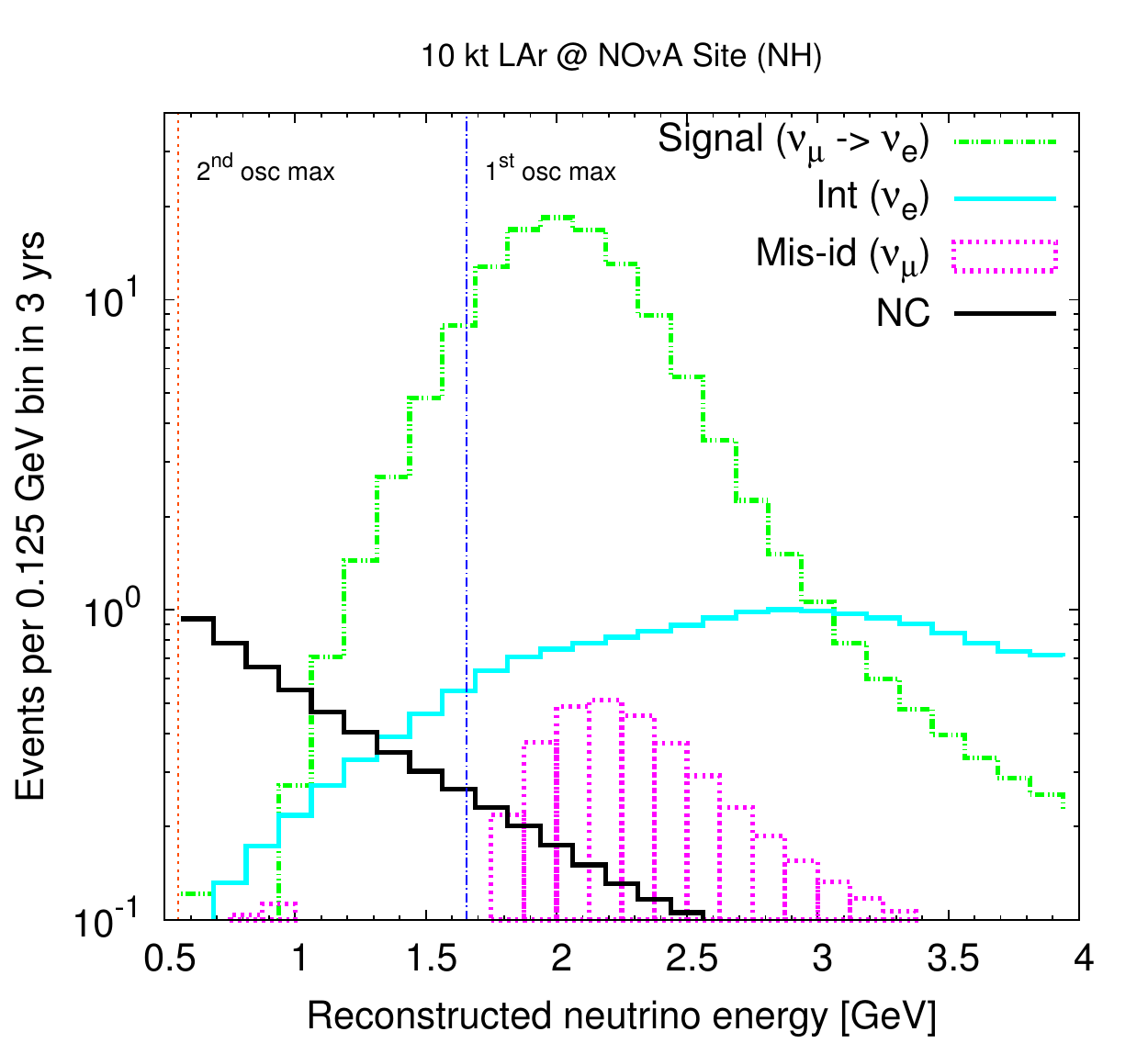}
\caption{{\footnotesize (colour online) Left panel portrays the expected signal and background event 
rates including the efficiency and background rejection capabilities 
in the $\nu_e$ appearance channel for $\sin^2 2 \theta_{13}$ = 0.101 
and $\dcp$ = 0$^{\circ}$, as a function of the reconstructed neutrino 
energy for the off-axis NO$\nu$A site with $14\,\mathrm{kt}$ TASD detector
running for 3 years exposed to $0.7\,\mathrm{MW}$ NuMI beam. Right panel 
shows the same for a $10\,\mathrm{kt}$ LArTPC detector running at off-axis 
NO$\nu$A site for 3 years exposed to $0.7\,\mathrm{MW}$ NuMI beam.
A normal hierarchy has been assumed. In all the panels, the blue dot-dashed 
and the orange dotted vertical lines display the locations of the first and
second oscillation maxima.}}
\label{fig:signal-bkg}
\end{figure}

The signal and different background event spectra are shown in figure ({\ref{fig:signal-bkg}}) 
for \nova (left panel) and for LArTPC (right panel). Even though the mass of the LArTPC 
detector is smaller, the event numbers in the two cases are comparable due to the higher signal acceptance. 
The NC background spectrum is shifted to lower energies due to the use of migration 
matrices and can easily be suppressed by an energy cut.

In Table-2, we list the number of signal and background events for the old \nova detector, 
new reoptimized \nova detector and for a $10\,\mathrm{kt}$ LArTPC, for both 
$\nu_\mu \rightarrow \nu_e$ appearance channel and $\nu_\mu \rightarrow \nu_\mu$ 
disappearance channel. In the calculation of the event numbers for the disappearance channel, 
we match the numbers with the plot given in NOvA website~\cite{nova_url}. 
The background events in the appearance channel
arise from three possible sources: (a) intrinsic $\nu_e/\bar{\nu}_e$
content of the beam, (b) mis-identified muons and (c) NC reactions.
In this study we have taken into account the $\nu_e$ contamination
in the $\bar{\nu}_\mu$ beam, which is a significant source of the
background.
In the disappearance channel backgrounds mainly arise from the 
$\bar{\nu}_\mu$ ($\nu_\mu$) contamination in $\nu_\mu$ ($\bar{\nu}_\mu$) 
beam with a small number coming from NC reactions. 
The signal events are shown for both neutrino and anti-neutrino 
runs and for both normal and inverted hierarchies.

We see from Table~\ref{tab2} that both the signal and background event rates are larger
for new \nova compared to old NO$\nu$A. The rise in the signal is only $50\%$ whereas the background
rises by a factor $3$. But the number of signal events increases by $30$ whereas the number of 
background events rises by $20$. Thus overall, there is a gain in signal relative to the background.
The background events are fixed and do not change with a change
in hierarchy or with a variation of $\dcp$, whereas the signal 
events do. Since we are interested in measuring these changes in 
the signal events with hierarchy and $\dcp$, it is more advantageous
to have a larger signal event sample, even at the expense of a 
larger background sample.

\begin{table}[H]
{\footnotesize
\begin{tabular}{||l||c|c||c|c||c|c||} 
\hline
\hline
\multicolumn{1}{||c||}{Channels}
&\multicolumn{2}{c||}{Old \nova(15 $\mathrm{kt}$)}
&\multicolumn{2}{c||}{New \nova(14 $\mathrm{kt}$)}
&\multicolumn{2}{c||}{LArTPC (10 $\mathrm{kt}$)}\\
\hline
\multicolumn{1}{||c||}{App.}
&\multicolumn{1}{c|}{Signal}
&\multicolumn{1}{c||}{Background}
&\multicolumn{1}{c|}{Signal}
&\multicolumn{1}{c||}{Background}
&\multicolumn{1}{c|}{Signal}
&\multicolumn{1}{c||}{Background}\\
\multicolumn{1}{||c||}{}
&\multicolumn{1}{c|}{{\tiny CC}}
&\multicolumn{1}{c||}{{\tiny (Int+Mis-ID+NC)}}
&\multicolumn{1}{c|}{{\tiny CC}}
&\multicolumn{1}{c||}{{\tiny (Int+Mis-ID+NC)}}
&\multicolumn{1}{c|}{{\tiny CC}}
&\multicolumn{1}{c||}{{\tiny (Int+Mis-ID+NC)}}\\
\hline
        $P_{\mu e}$(NH)
         &{\bf62} &6+1+4={\bf11} 
         &{\bf92} &8+5+19={\bf32} 
         &{\bf123} &18+5+7={\bf30} \\
         $P_{\mu e}$(IH)
         &{\bf36} &6+1+4={\bf11} 
         &{\bf54} &8+5+19={\bf32} 
         &{\bf72} &19+5+7={\bf31} \\
         $P_{\bar{\mu} \bar{e}}$(NH)
         &{\bf26} &6+$<$1+6={\bf12} 
         &{\bf30} &5+$<$1+10={\bf15} 
         &{\bf28} &17+2+2={\bf21} \\
         $P_{\bar{\mu} \bar{e}}$(IH)
         &{\bf34} &5+$<$1+6={\bf11} 
         &{\bf38} &5+$<$1+10={\bf15} 
         &{\bf36} &14+2+2={\bf18} \\
\hline
\multicolumn{1}{||c||}{Disapp.}
&\multicolumn{1}{c|}{Signal}
&\multicolumn{1}{c||}{Background}
&\multicolumn{1}{c|}{Signal}
&\multicolumn{1}{c||}{Background
}&\multicolumn{1}{c|}{Signal}
&\multicolumn{1}{c||}{Background}\\
\multicolumn{1}{||c||}{}
&\multicolumn{1}{c|}{{\tiny CCQE}}
&\multicolumn{1}{c||}{{\tiny (NC only)}}
&\multicolumn{1}{c|}{{\tiny CCQE}}
&\multicolumn{1}{c||}{{\tiny (NC+Wrong-Sign muon)}}
&\multicolumn{1}{c|}{{\tiny CCQE}}
&\multicolumn{1}{c||}{{\tiny (NC+Wrong-Sign muon)}}\\
\hline
         $P_{\mu \mu}$(NH)
         &{\bf173} &{\bf2} 
         &{\bf134} &1+6={\bf7} 
         &{\bf403} &7+20={\bf27} \\
         $P_{\mu \mu}$(IH)
         &{\bf173} &{\bf2} 
         &{\bf134} &1+6={\bf7} 
         &{\bf402} &7+20={\bf27} \\
         $P_{\bar{\mu} \bar{\mu}}$(NH)
         &{\bf102} &{\bf1} 
         &{\bf43} &$<$1+18={\bf18} 
         &{\bf136} &2+54={\bf56} \\
         $P_{\bar{\mu} \bar{\mu}}$(IH)
         &{\bf103} &{\bf1} 
         &{\bf43} &$<$1+18={\bf18} 
         &{\bf137} &2+54={\bf56} \\
\hline
\hline
\end{tabular}
}
\caption{{\footnotesize Total number of signal and background events for
old NO$\nu$A, new \nova and LArTPC, in both appearance $(\nu_\mu 
\rightarrow \nu_e)$ and disappearance $(\nu_\mu \rightarrow
\nu_\mu)$ modes. The backgrounds are subdivided into three 
parts: (a) intrinsic beam $\nu_e/\bar{\nu}_e$ (Int.), (b)
mis-identified muons (Mis-ID) and (c) single $\pi^0$ events
from neutral current interactions (NC). In the disappearance
mode, wrong-sign muon indicates the background coming from
$\bar{\nu}_\mu~(\nu_\mu)$ contamination in $\nu_\mu (\bar{\nu}_\mu)$
beam.}}
\label{tab2}
\end{table}

\begin{figure}[H]
\centering
\includegraphics[width=0.49\textwidth]{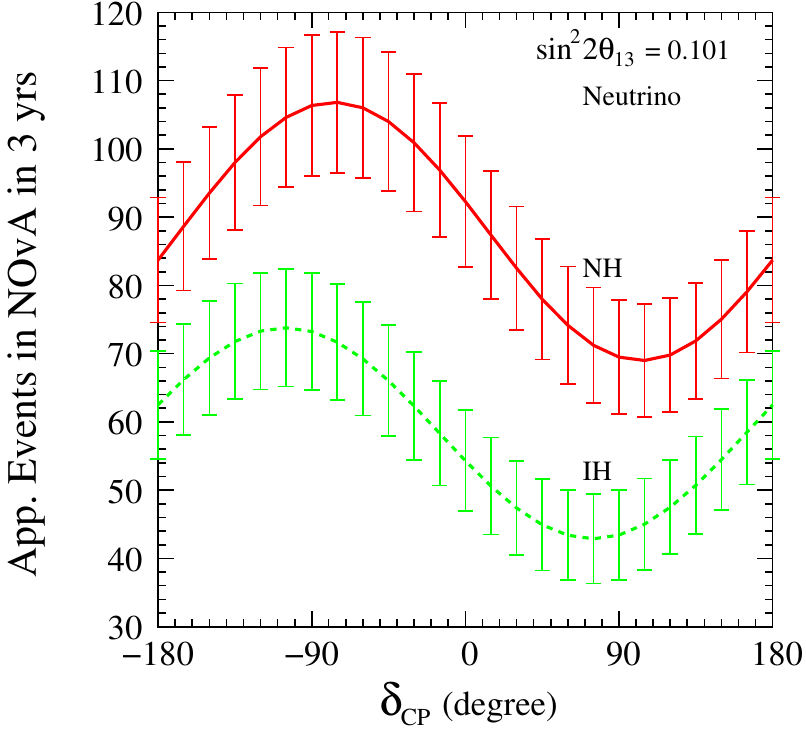}
\includegraphics[width=0.49\textwidth]{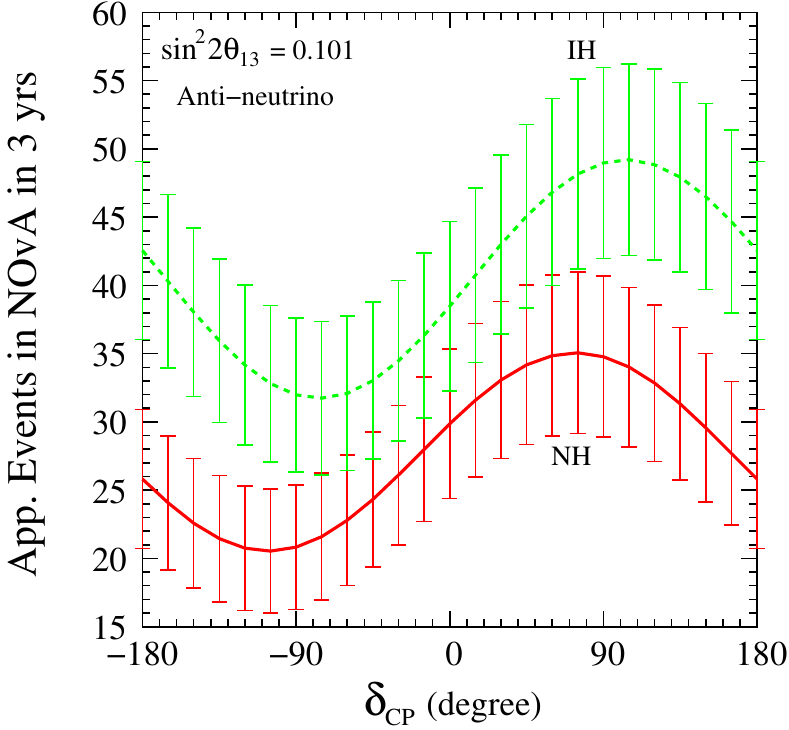}
\caption{{\footnotesize (colour online) Events vs. $\dcp$. Left panel is for $\nue$ appearance 
and right panel is for $\anue$ appearance. The error bars shown are statistical. 
Please note the difference in the scale of y-axis in the two panels.}}
\label{fig:eventsvsdcp}
\end{figure}

In figure (\ref{fig:eventsvsdcp}), we plot the signal event numbers
(along with statistical error bars) as a function of $\dcp$ 
for new NO$\nu$A, both for NH and IH. The left panel shows
the event numbers for three years of running in the neutrino mode 
and the right panel depicts the event numbers for three years in 
the anti-neutrino mode. We see that for the combination 
NH and LHP ($-180^\circ \leq \dcp \leq 0$), the neutrino event numbers
are much greater than the IH event numbers for any $\dcp$.
Similarly for the combination IH and UHP ($0 \leq \dcp \leq 180^\circ$),
the event numbers are much lower than NH event numbers for any
$\dcp$. Hence \nova has good hierarchy discrimination for these
two combinations. For the other two combinations, it is possible
to have two solutions with different hierarchies for the same
event numbers. This feature occurs, not only for the total event
numbers but also for the spectrum. Thus, for unfavourable $\dcp$
half-planes, hierarchy determination is not possible with NO$\nu$A.

\section{Numerical Simulation}

We use the minimization of $\Delta\chi^2$ to estimate the hierarchy and CP violation sensitivities.
For hierarchy sensitivity, we first assume NH to be the true hierarchy and we choose a true value of $\dcp$. 
We compute the NH event spectrum for these assumptions and the above
true values of neutrino parameters and label it to be {\it data}. Then we 
compute a theoretical event spectrum assuming IH and varying the
test values of neutrino parameters within their $\pm 2\sigma$ ranges
and $\dcp$ in the full allowed range $(-180^\circ, 180^\circ)$.
In doing this marginalization, we impose Gaussian priors on the measured
neutrino parameters. We compute a $\Delta\chi^2$ between the event
spectra of the data and the theory and demand that its minimum value
$\Delta\chi^2_{\mathrm min} \geqslant 2.71~(3.84)$ for a 90\% (95\%) C.L. hint of hierarchy.
If this condition is satisfied, then the hierarchy can be determined
at the appropriate confidence level, for the assumption of hierarchy being NH and $\dcp$ 
being the assumed true value. The calculation is repeated for various other assumed true
values of $\dcp$. If $\Delta\chi^2_{\mathrm min} \geqslant 2.71~(3.84)$ 
for all values of true $\dcp$, then we can say that the normal hierarchy 
can be established for all possible values of neutrino parameters.
The whole calculation then has to be repeated for the case of 
true hierarchy being IH. If $\Delta\chi^2_{\mathrm min} \geqslant 2.71~(3.84)$
for all true values of $\dcp$, then the hierarchy can be determined for all possible values of neutrino parameters.

In computing the above spectra, it should be noted that the values of $\Delta m^2_{31}$ for NH and for IH are different. 
They are to be calculated from the expression in eq.~(\ref{parkedef}) with $\Delta m^2_{\mathrm{eff}}$ positive for NH 
and negative for IH. Since $\Delta m^2_{21}$ is always positive, this leads to different
magnitudes for $\Delta m^2_{31}$ for NH and for IH. This difference must be taken into account while calculating 
the NH and IH spectra. Otherwise, there will be a spurious hierarchy sensitivity in the 
disappearance channel.

As mentioned in the introduction, for a given experiment,
we will get two degenerate solutions: one with the correct
hierarchy and true $\dcp$ and one with the wrong hierarchy
and a wrong $\dcp$. It was shown in~\cite{menaparke} that 
two experiments with different baselines, with flux peaking
at the first oscillation maximum, can pick the correct hierarchy
and $\dcp$ because the wrong hierarchy solutions for different
experiments occur for different values of $\dcp$. But, the
statistics of each of the two experiments have to be large enough
for a clean separation. We will see in the next section that
the presently planned runs of T2K and \nova are not enough
for this separation to occur for the full range of $\dcp$.
Therefore, for the existence of CP violation, we 
pose the following question: for what values of true $\dcp$
can \nova and T2K establish that the CP phase differs from 
0 or $180^\circ$, {\it independently of the hierarchy}? 
To answer this question, we compute the data spectrum say for NH and 
for a true value of $\dcp$, as described above. We compute
the theoretical spectrum for the following four combinations 
of hierarchy and CP conservation (NH, $\dcp=0$), (NH, $\dcp=180^\circ$),
(IH, $\dcp=0$) and (IH, $\dcp=180^\circ$). We compute the $\Delta \chi^2$
between the data and each of these four combinations and choose 
its minimum value. If $\Delta \chi^2_{\mathrm min} \geqslant 2.71~(3.84)$,
then CP violation is established at $90\%$ ($95\%$) C.L.
for true hierarchy being NH and for $\dcp$ being the assumed true
value. We repeat the calculation for the full range of $\dcp$
and determine the values of true $\dcp$ for which
CP violation can be established. As in the case of hierarchy determination, 
these calculations are again repeated for the case where IH is the true hierarchy.

It must be mentioned that this method only establishes that the CP conservation is ruled out. 
But it is possible that the data may not be able to determine the hierarchy. In
such a situation, the data must be analyzed under the assumption
that either of the two hierarchies is correct. If the data is analyzed 
with the assumption of the right hierarchy, we get a region of
allowed $\dcp$ surrounding the true $\dcp$. If the data is 
analyzed under the assumption of the wrong hierarchy, then the
allowed $\dcp$ region may not include the true $\dcp$. The above
procedure only requires that, for each of the hierarchies, the
allowed $\dcp$ region should not include the CP conserving cases
$\dcp = 0$ and $180^\circ$. If this condition is satisfied, the
CP violation in neutrino sector is established independent of the
hierarchy. We will not be able to know the true value of $\dcp$
without determining the hierarchy~\cite{novat2k}. But proving the existence of
of leptonic CP violation will be an important step forward in 
our understanding of the leptonic masses and mixings. 

\section{Results}

In this section, we describe the capabilities of the experimental setups considered
in this work, for the determination of mass hierarchy and CP violation.

\subsection{Mass Hierarchy Discrimination}

\begin{figure}[H]
\centering
\includegraphics[width=0.49\textwidth]{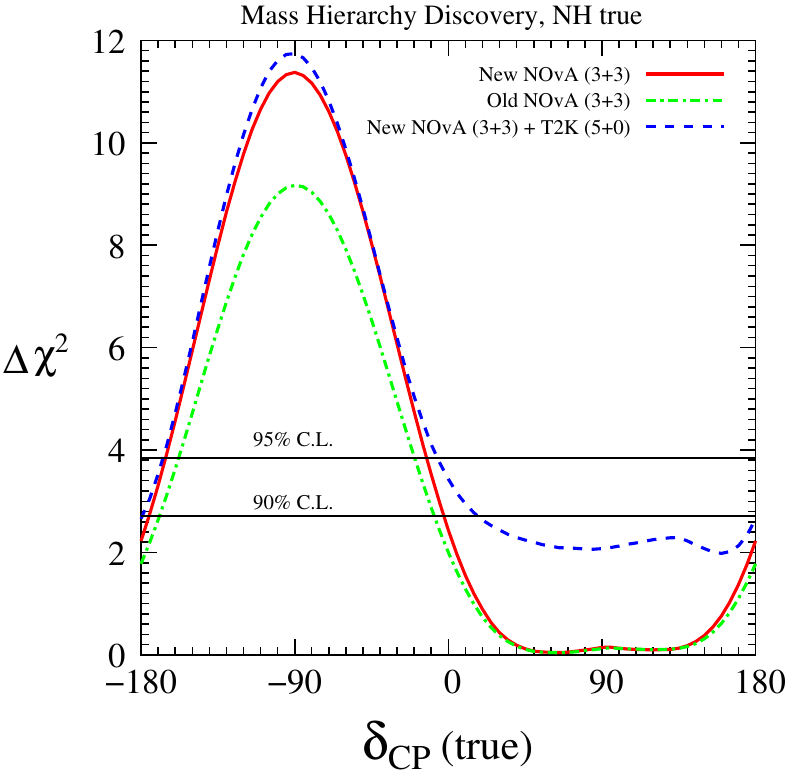}
\includegraphics[width=0.49\textwidth]{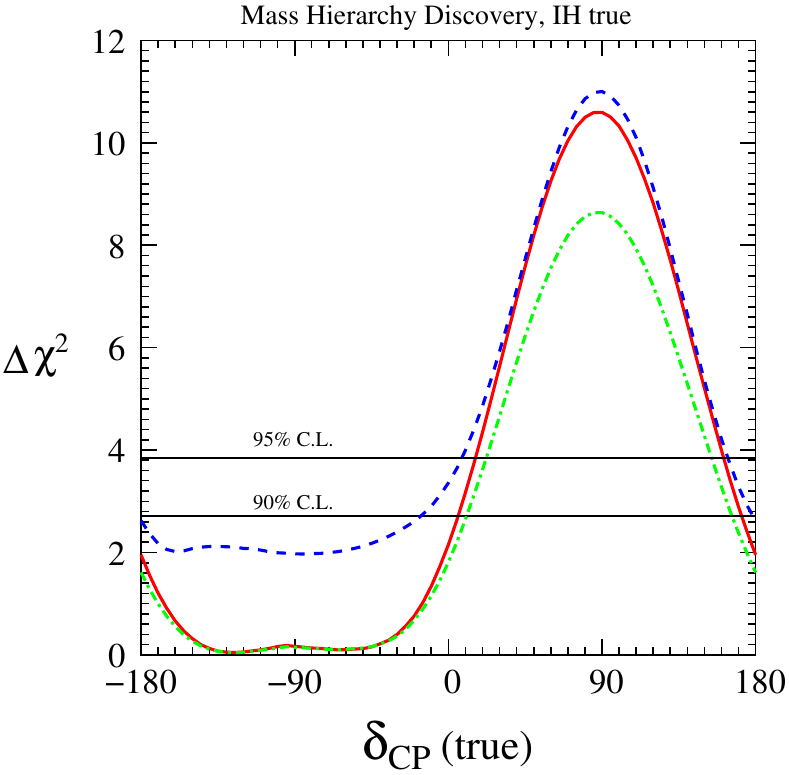}
\caption{{\footnotesize (colour online) Mass hierarchy discovery as a function of true value of $\dcp$. 
Left (right) panel is for NH (IH) as true hierarchy.}}
\label{fig:mhdiscovery}
\end{figure}

In figure (\ref{fig:mhdiscovery}), we plot the hierarchy discrimination
sensitivity of the old NO$\nu$A, the new \nova and the combined sensitivity
of new \nova and T2K, as a function of the true value of $\dcp$, in
the left (right) panel for NH (IH) as the true hierarchy. We see that
the wrong hierarchy can be ruled out very effectively for $\dcp$
in the favourable half-plane, which is LHP (UHP)  
for NH (IH). 

The new event selection criteria of \nova make the experiment 
even more effective in ruling out the wrong hierarchy for $\dcp$ in
the favourable half-plane. In the unfavourable half-plane, both the
old and the new criteria are equally ineffective. However, the addition
of T2K data improves the situation significantly and $\Delta \chi^2$ 
increases from 0 to $\geq 2$ for all the true values of $\dcp$,
thus making it possible to get a $90\%$ C.L. hint of hierarchy with 
some additional data. We have checked that a further increment in the 
exposure of T2K or addition of antineutrino data from T2K 
does not improve the hierarchy sensitivity much.

\begin{figure}[H]
\centering
\includegraphics[width=0.49\textwidth]{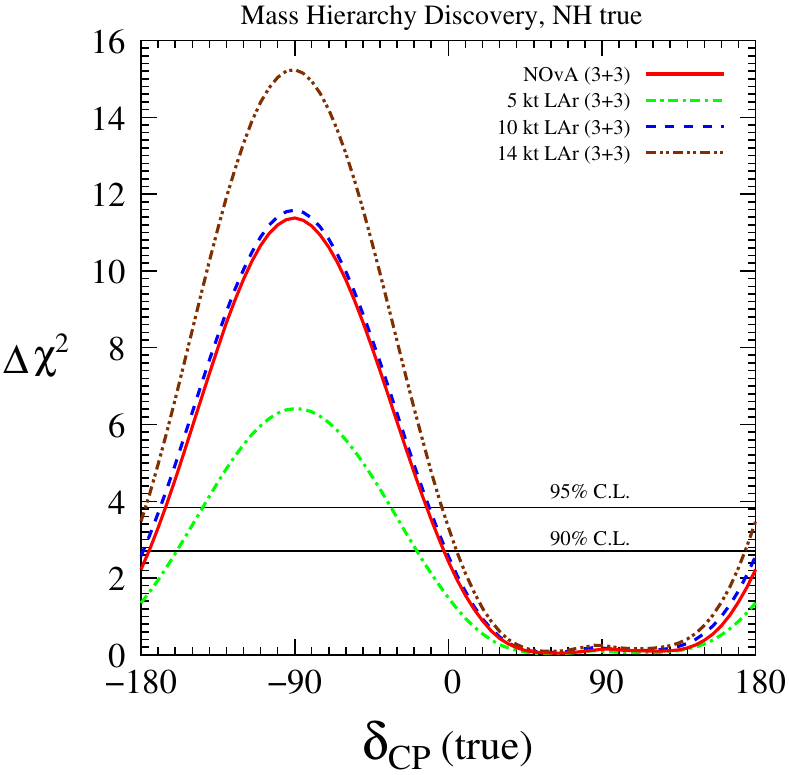}
\includegraphics[width=0.49\textwidth]{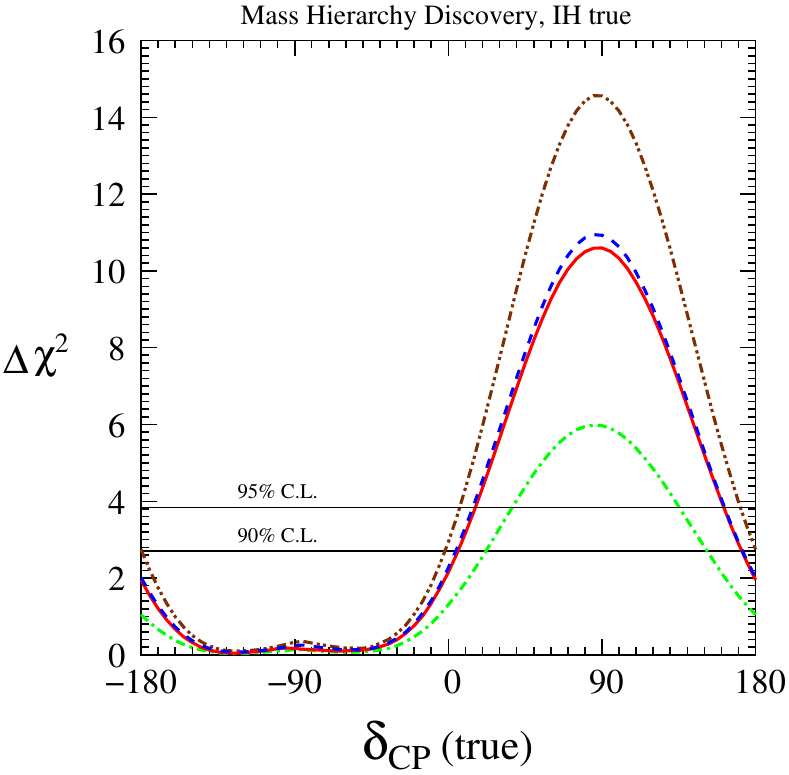}
\caption{{\footnotesize (colour online) Mass hierarchy discovery as a function of true value of $\dcp$, 
for different LArTPC detector masses and new NO$\nu$A. Left (right) panel is for NH (IH) as true hierarchy.}}
\label{fig:mhdiscovery-LAr-NOvA}
\end{figure}

In figure (\ref{fig:mhdiscovery-LAr-NOvA}), we plot the mass hierarchy
discrimination capability of \nova and a stand alone LArTPC with
three possible masses: 5 $\mathrm{kt}$, 10 $\mathrm{kt}$ and 14 $\mathrm{kt}$. 
Before the reoptimization of \nova event selection criteria, it was 
argued that a 5 $\mathrm{kt}$ LArTPC has the same capability 
as of the 15 $\mathrm{kt}$ TASD detector, because the signal acceptance
of the former was three times that of the latter~\cite{jennylbno}. But with the 
new event selection criteria the performance of \nova has dramatically 
improved. We see from the figure (\ref{fig:mhdiscovery-LAr-NOvA}) 
that only a LArTPC of mass 10 $\mathrm{kt}$ can be as effective
as NO$\nu$A. But, once again, it must be noted that none of the detectors
are effective if the true value of $\dcp$ is in the unfavourable half-plane.

\begin{figure}[H]
\centering
\includegraphics[width=0.49\textwidth]{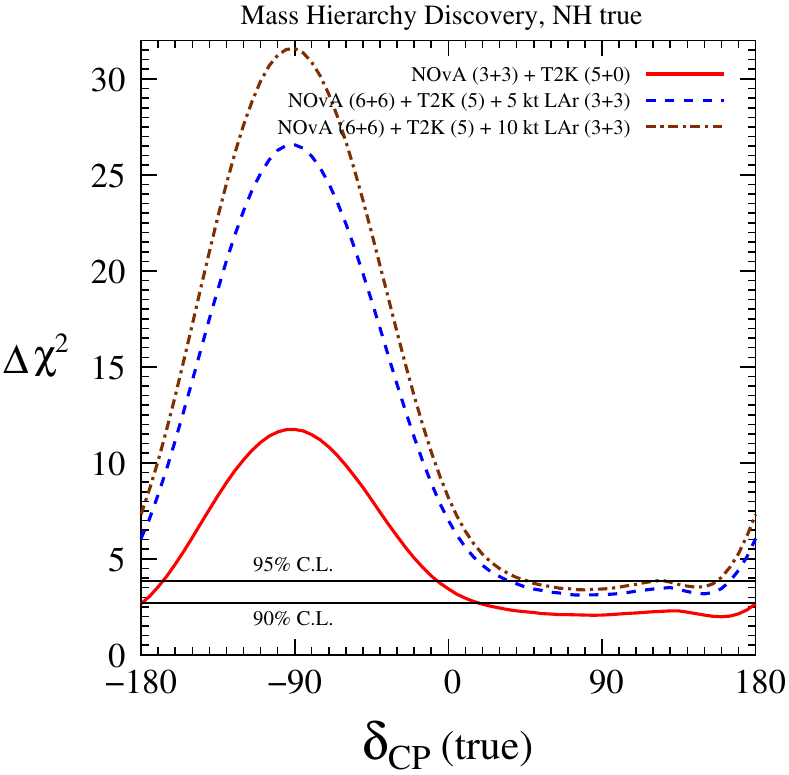}
\includegraphics[width=0.49\textwidth]{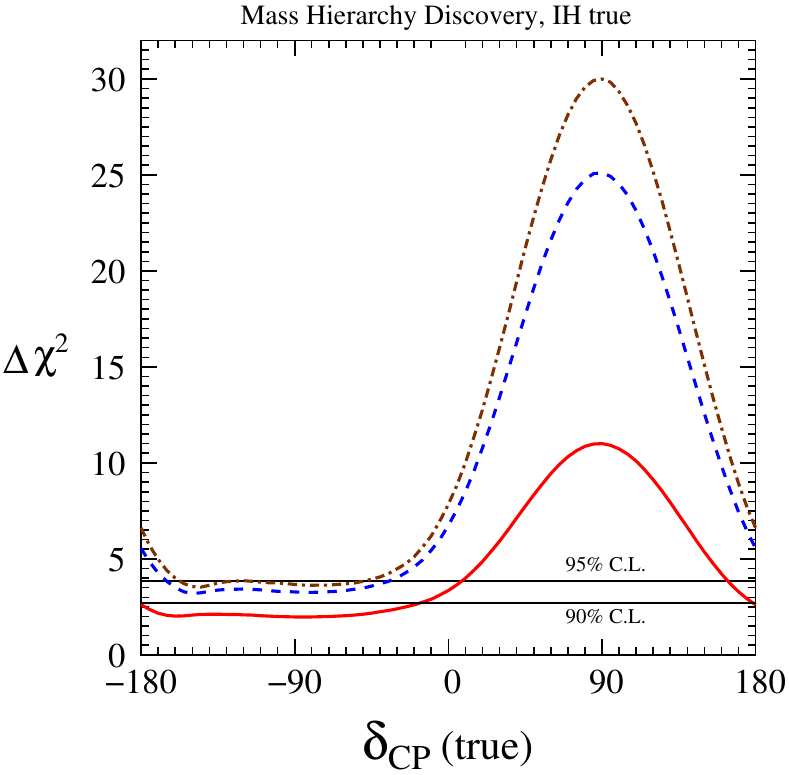}
\caption{{\footnotesize (colour online) Mass hierarchy discovery as a function of
true value of $\dcp$. Combined performance of new NO$\nu$A, 
T2K and LArTPC. Left (right) panel is for NH (IH) as true hierarchy.}}
\label{fig:mhdiscoverycomb}
\end{figure}

Finally, in figure (\ref{fig:mhdiscoverycomb}) we plot the combined hierarchy
discovery sensitivity of NO$\nu$A, T2K and a LArTPC (of mass 5 $\mathrm{kt}$ 
and 10 $\mathrm{kt}$) and compare it with the hierarchy sensitivity of combined 
\nova and T2K data. We find that the addition of even a 5 $\mathrm{kt}$ 
LArTPC leads to a very significant improvement in the sensitivity when
values of true $\dcp$ are in the unfavourable half-plane. With the 
addition of a 10 $\mathrm{kt}$ LArTPC, we see that close 
to 95$\%$ C.L. hierarchy discrimination becomes possible for all the values
of $\dcp$.

\subsection{CP Violation Discovery}

\begin{figure}[H]
\centering
\includegraphics[width=0.49\textwidth]{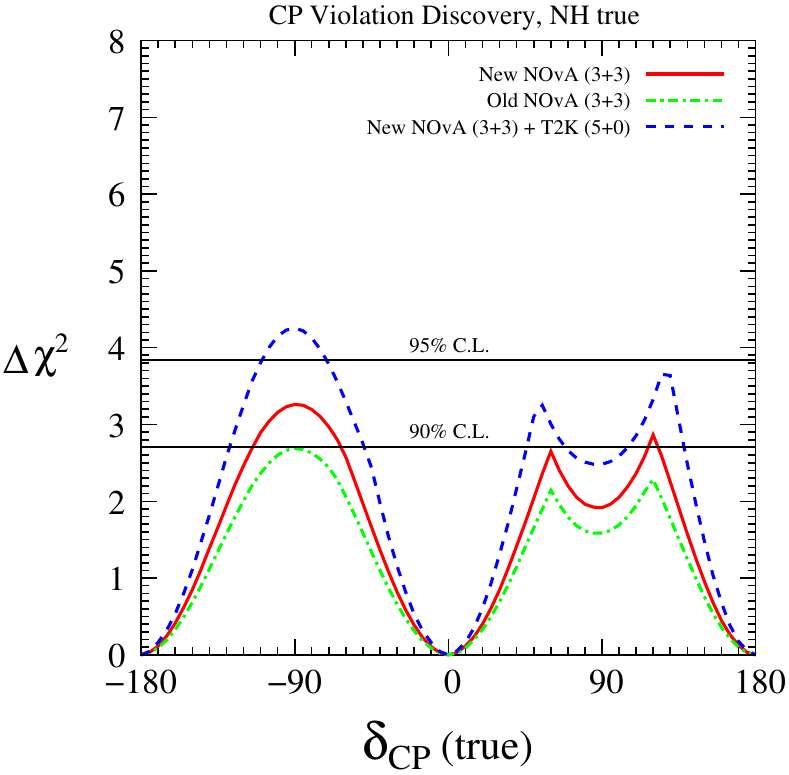}
\includegraphics[width=0.49\textwidth]{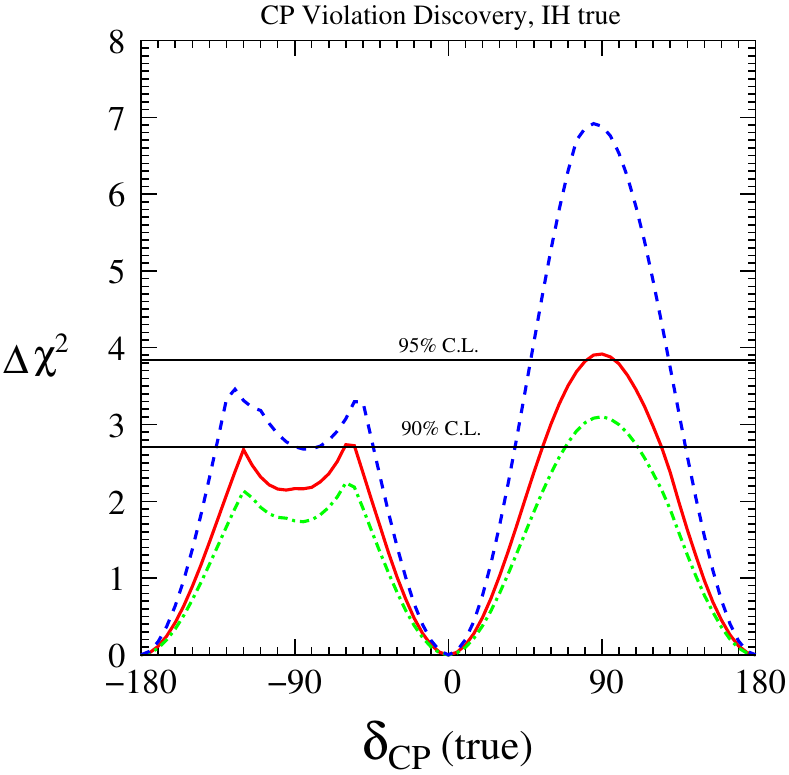}
\caption{{\footnotesize (colour online) CP violation discovery as a function of 
true value of $\dcp$. Left (right) panel is for NH (IH) as true hierarchy.}}
\label{fig:cpvdiscovery}
\end{figure}

Reoptimization of the event selection criteria of \nova has
the most dramatic effect on the CP violation discovery potential
of the experiment. In figure (\ref{fig:cpvdiscovery}), we plot the
sensitivity to rule out CP conservation scenarios, as a function of true
$\dcp$ in the left (right) panel for NH (IH) being the true 
hierarchy. We notice that, while in the case of old NO$\nu$A there
is no CP violation sensitivity at all at 90\% C.L., there is such a sensitivity
in new NO$\nu$A, for about one third fraction of the favourable 
half-plane. Addition of T2K data leads to CP violation sensitivity for
about half the region in both favourable half planes at 90\% C.L..
It can be shown that, T2K by itself, has no CP violation sensitivity.
But, the synergistic combination of \nova and T2K leads to much 
better CP violation sensitivity compared to the individual capabilities.

\begin{figure}[H]
\centering
\includegraphics[width=0.49\textwidth]{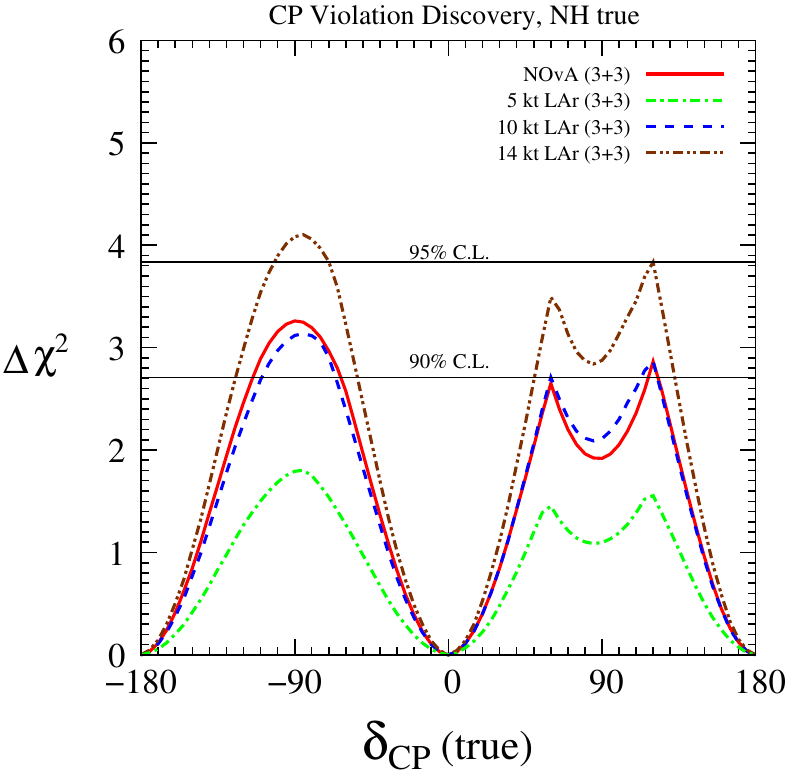}
\includegraphics[width=0.49\textwidth]{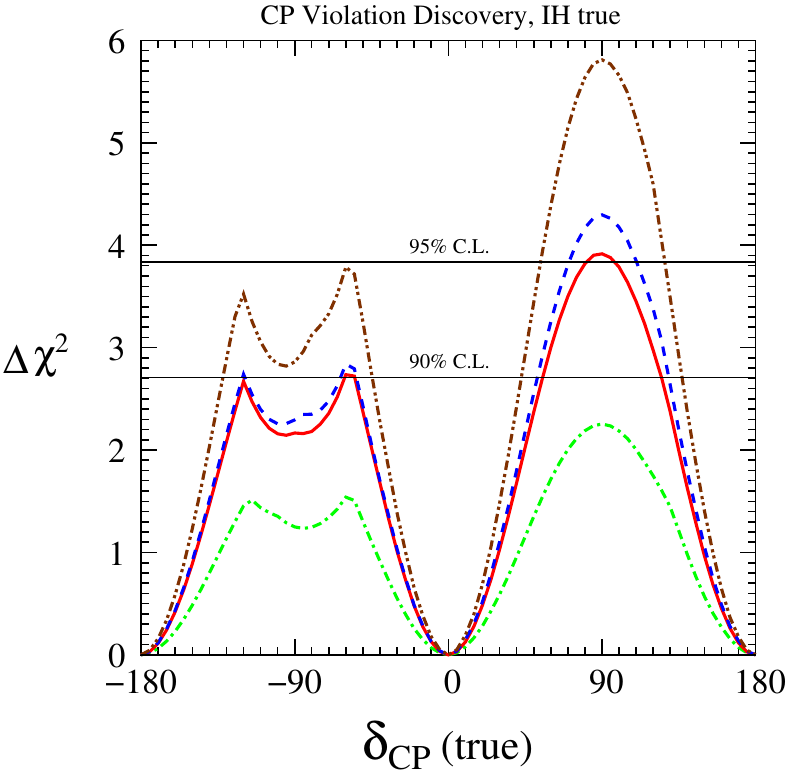}
\caption{{\footnotesize (colour online) CP violation discovery as a function of true value of $\dcp$, 
for different LArTPC detector masses and new NO$\nu$A. 
Left (right) panel is for NH (IH) as true hierarchy.}}
\label{fig:cpvdiscoverylarnova}
\end{figure}

In figure (\ref{fig:cpvdiscoverylarnova}), we present a comparison of
new \nova with stand alone LArTPC of different masses.
As in the case of the mass hierarchy discrimination, the
performance of a 10 $\mathrm{kt}$ LArTPC detector is closest to the
performance of the new NO$\nu$A.

\begin{figure}[H]
\centering
\includegraphics[width=0.49\textwidth]{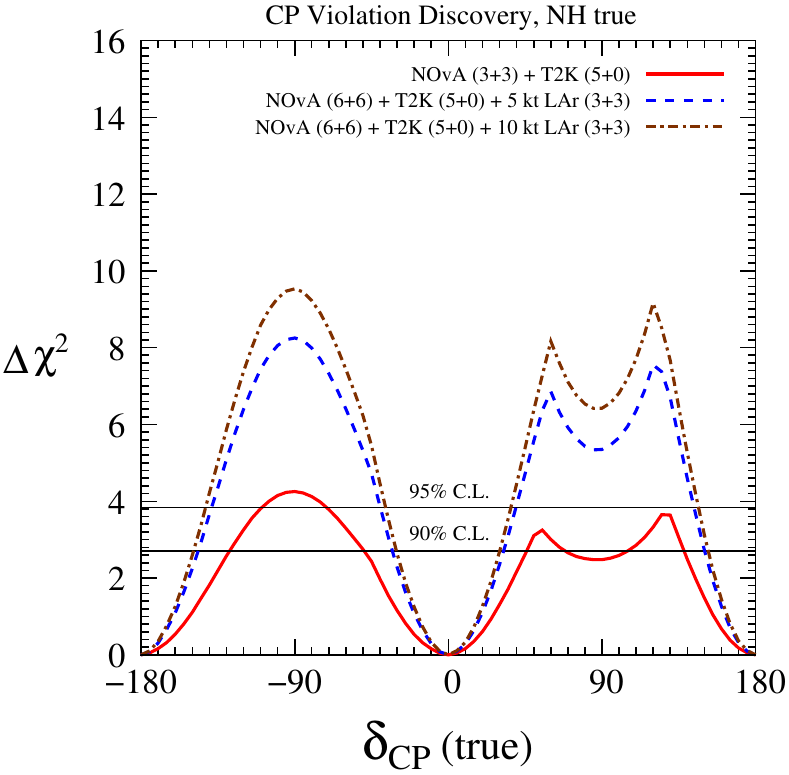}
\includegraphics[width=0.49\textwidth]{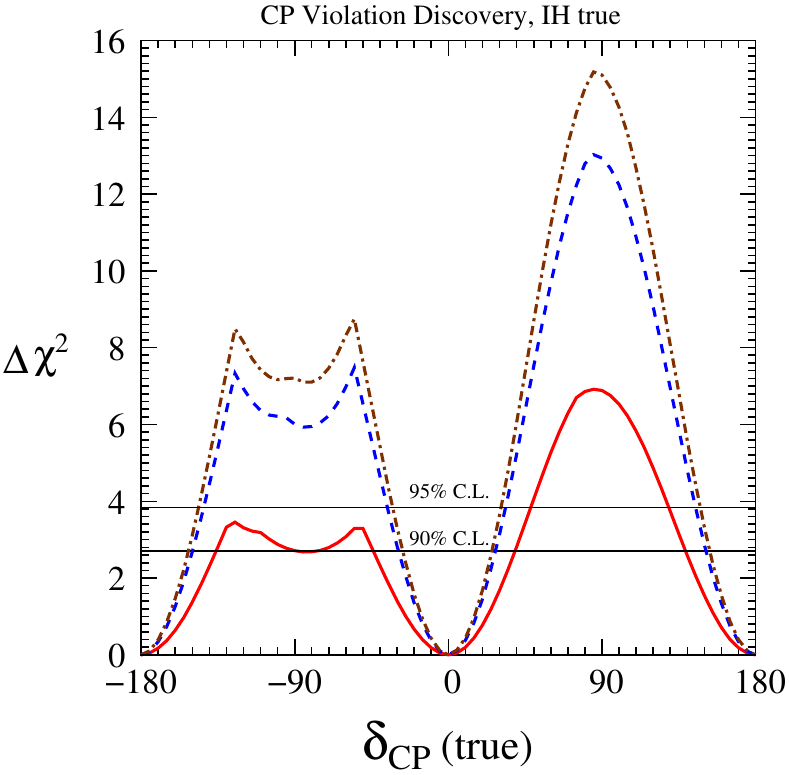}
\caption{{\footnotesize (colour online) CP violation discovery as a function of 
true value of $\dcp$. Combined performance of new NO$\nu$A, 
T2K and LArTPC. Left (right) panel is for NH (IH) as true hierarchy.}}
\label{fig:cpvdiscoverycomb}
\end{figure}

In figure (\ref{fig:cpvdiscoverycomb}), we plot the CP violation discovery
potential of the combined data from new \nova with a 
$(6 \nu + 6 \bar{\nu})$ run, LArTPC of mass 5 $\mathrm{kt}$ and 10 $\mathrm{kt}$ 
with a $(3 \nu + 3 \bar{\nu})$ run and T2K along with that of combined new \nova with a  
$(3 \nu + 3 \bar{\nu})$ run and T2K data. The net effect of new \nova with $(6 \nu + 6 \bar{\nu})$ run plus a
10 $\mathrm{kt}$ LArTPC with a $(3 \nu + 3 \bar{\nu})$ is equivalent to tripling the data of
\nova with $(3 \nu + 3 \bar{\nu})$ run. We note that the addition of the LArTPC leads to a very
significant improvement in the range of true $\dcp$ for which 
CP conservation can be ruled out at 95\% C.L.. This range goes up from about half of the allowed values 
to about 60\% of the allowed values. 

A summary of our results is given in Table~\ref{tab:compare}. For different 
combinations of experiments and different exposures, we have shown the fraction 
of $\dcp$ values for which mass hierarchy can be determined/CP violation can be 
detected.

\begin{table}[H]
\begin{center}
{\footnotesize
\begin{tabular}{|c|c|c||c|c|} \hline\hline
\multirow{4}{*}{Setups} & \multicolumn{4}{c|}{{\rule[0mm]{0mm}{6mm}Fraction of $\dcpt$}}
\cr\cline{2-5}
& \multicolumn{2}{c||}{{\rule[0mm]{0mm}{4mm}{\bf MH}}} & \multicolumn{2}{c|}{\rule[0mm]{0mm}{4mm}{{\bf CPV}}} 
\cr
\cline{2-5}
& NH true & IH true & NH true & IH true \cr
\hline
 NO$\nu$A (3+3) & 0.48 (0.43) & 0.46 (0.41) & 0.16 (0) & 0.21(0.04) \cr
\hline
 NO$\nu$A (3+3) + T2K (5+0) & 0.55 (0.45) & 0.54 (0.43) & 0.38 (0.11) & 0.49 (0.23) \cr
\hline
 NO$\nu$A (6+6) + T2K (5+0) + 5 kt LArTPC (3+3) & 1 (0.64) & 1 (0.64) & 0.64 (0.56) & 0.68 (0.61) \cr
\hline
 NO$\nu$A (6+6) + T2K (5+0) + 10 kt LArTPC (3+3) & 1 (0.71) & 1 (0.73) & 0.67 (0.60) & 0.71 (0.64) \cr
\hline\hline
\end{tabular}
}
\caption{{\footnotesize Fractions of true values of $\dcp$ for which a discovery is possible for MH and CPV. 
The numbers without (with) parentheses correspond to 90\% (95\%) C.L. 
The results are shown for both NH and IH as true hierarchy.}}
\label{tab:compare}
\end{center}
\end{table}

\section{Summary and Conclusions}

In light of the recent measurements indicating a moderately large value 
of $\sin^2 2\theta_{13}$, the \nova collaboration has revisited their 
background rejection cuts. The
relaxed acceptance cuts have resulted in a higher number of signal events.
Hence, the variations induced by the change in hierarchy or a change in $\dcp$
are magnified and become more easy to detect.

In this work, we have studied the physics reach of the reoptimized \nova (in conjunction 
with T2K) to determine the mass hierarchy and CP violation. With the presently
planned runs of these two experiments, the above goals can be achieved for less
than half the allowed values of $\dcp$. There is 
considerable enthusiasm in the \nova and LBNE collaborations for an additional 
small liquid argon detector detector at the Ash River site. In this study we 
have also considered the possibility of such a module. 

For favourable values of $\dcp$, \nova in its original configuration can 
determine the mass hierarchy by itself. The reoptimized new \nova allows 
us to determine the hierarchy with greater confidence in the favourable 
half-plane of $\dcp$. However, in the unfavourable half-plane, the 
sensitivity is negligible. Addition of data from T2K helps to break the 
parameter degeneracy and increases the $\Delta\chi^2$ significantly in the 
unfavourable half-plane. But the currently planned runs of \nova and
T2K are not sufficient to raise the $\Delta\chi^2$ above 2.71 and give
us a 90$\%$ hint of hierarchy for $\dcp$ in the 
unfavourable half-plane. 

A 10 kt LArTPC is found to be equivalent to new \nova in its ability to 
exclude the wrong hierarchy. Thus, addition of such a module to the existing 
\nova detector will help increase the statistical significance of the 
experimental data. We find that for increased exposure for NO$\nu$A, the 
combination of data from NO$\nu$A, LArTPC and T2K (nominal) can determine 
hierarchy at almost $2\sigma$ level for all the values of $\dcp$. We have checked 
that increasing the exposure of T2K or adding antineutrino data from T2K 
does not improve the results much. 

Discovering CP violation is more difficult than determining the hierarchy. 
\nova by itself can discover CP violation only for a small fraction of the 
favourable half-plane, and only at $90\%$ C.L. Addition of data from T2K
causes a remarkable increase, by a factor of 2.4, in the range of $\dcp$
for which CP violation can be established. This includes a significant part of
the unfavourable half-plane also. Once again, we find that a 10 kt 
LArTPC has capabilities similar to NO$\nu$A. The combination of additional 
data from the LArTPC detector with boosted \nova and nominal T2K greatly improves 
the ability. For around 60\% of the entire $\dcp$ range, CP violation can 
be discovered at 95\% confidence level. 
 
In conclusion, we find that after reoptimizing \nova for large $\theta_{13}$, 
the sensitivity of the experiment to mass hierarchy and CP violation is 
increased. Adding data from T2K breaks the hierarchy-$\dcp$ degeneracy but 
is not enough to determine hierarchy for unfavourable $\dcp$ values. Additional 
data from a LArTPC will lead to a significant boost in both hierarchy and 
CP violation sensitivities when combined with \nova and T2K. 

\subsubsection*{Acknowledgments}

S.K.A. would like to thank Ryan Patterson and Geralyn Zeller for useful discussions.
S.K.A. also acknowledges the support from the European Union under the European 
Commission FP7 Research Infrastructure Design Studies
EUROnu (Grant Agreement No.  212372 FP7-INFRA-2007-1), LAGUNA (Grant Agreement No. 212343 FP7-INFRA-2007-1) and the project
Consolider-Ingenio CUP. S.K.R. would like to thank Jenny Thomas for useful discussions.

\begin{appendix}

\section{Performance with conservative choices of central values}
\label{appendix1}

Here we consider how the sensitivities of various setups will change
if the conservative input values of neutrino mixing angles, mentioned
in section 3, are used. 
In table~\ref{tab:bestfit}, we list the values  
considered in the main paper (called Best-fit 1) and the 
conservative values considered in this appendix (called Best-fit 2). 

\begin{table}[H]
\begin{center}
{\footnotesize
\begin{tabular}{|c|c|c|} \hline \hline
 & Best-fit 1 & Best-fit 2 \\ \hline
$\sin^2 2 \theta_{13}$ & $0.101$ & $0.089$ \\ \hline
$\sin^2 \theta_{23}$ & $0.5$ & $0.413$ \\ \hline
\end{tabular}
}
\caption{{\footnotesize The values of $\sin^22\ty$ and $\sin^2\tz$ taken in Best-fit 1 and
Best-fit 2. The values of other oscillation parameters are the same for the two cases considered.}}
\label{tab:bestfit}
\end{center}
\end{table} 

\begin{figure}[H]
\centering
\includegraphics[width=0.49\textwidth]{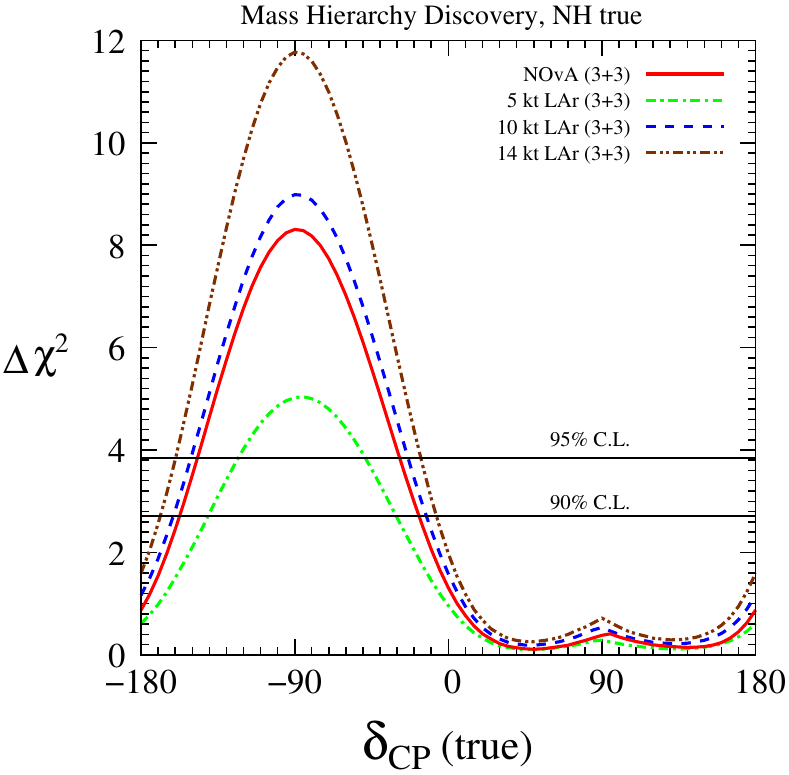}
\includegraphics[width=0.49\textwidth]{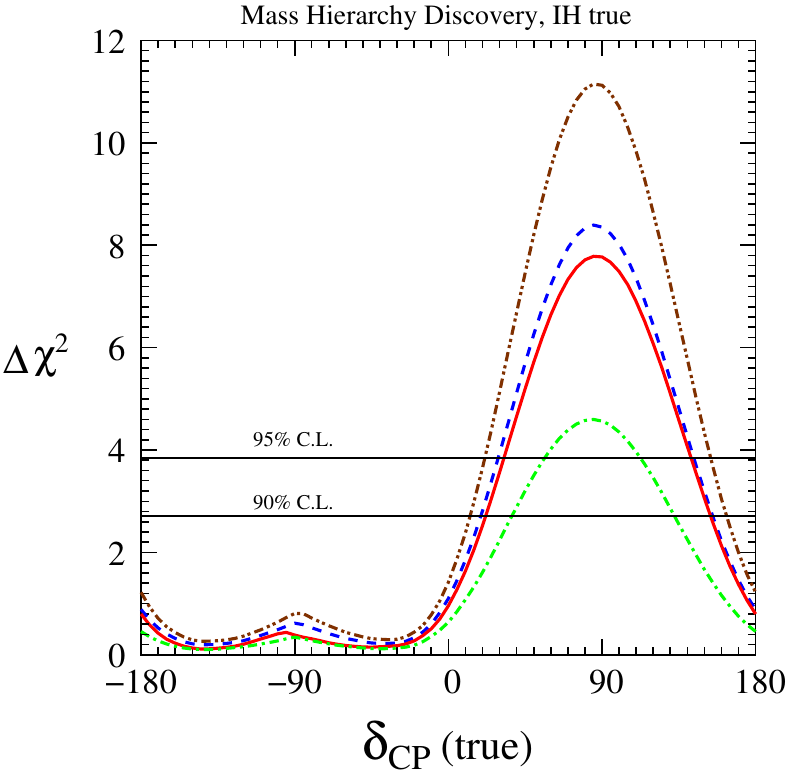}
\caption{{\footnotesize (colour online) Mass Hierarchy discovery as a 
function of 
true value of $\dcp$ for the conservative values of input parameters,
of new NO$\nu$A and of LArTPC of various masses. 
Left (right) panel is for NH (IH) as true hierarchy.}}
\label{fig:mhdiscovery-LAr-NOvA2}
\end{figure}

\begin{figure}[H]
\centering
\includegraphics[width=0.49\textwidth]{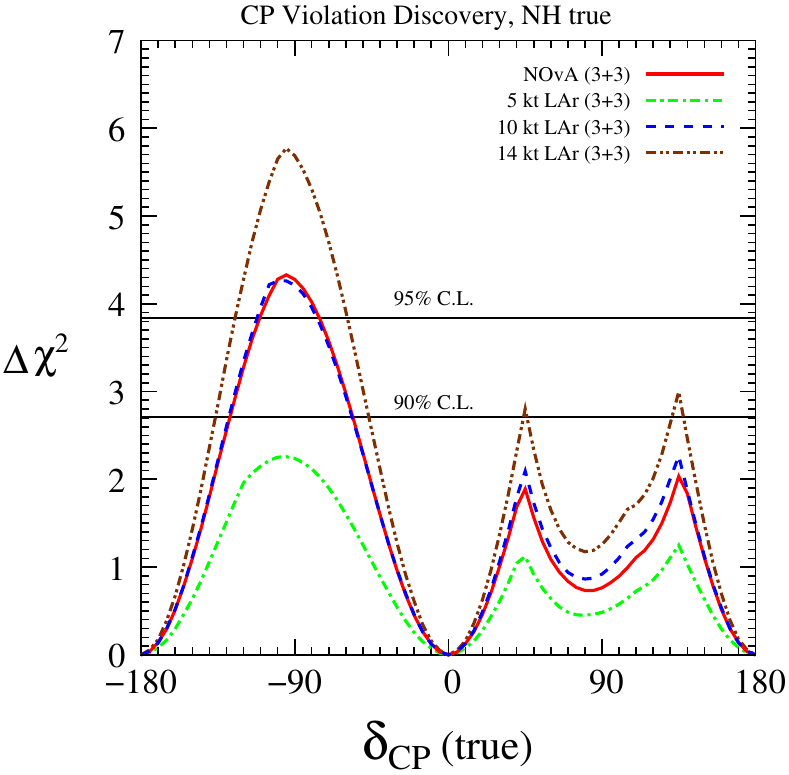}
\includegraphics[width=0.49\textwidth]{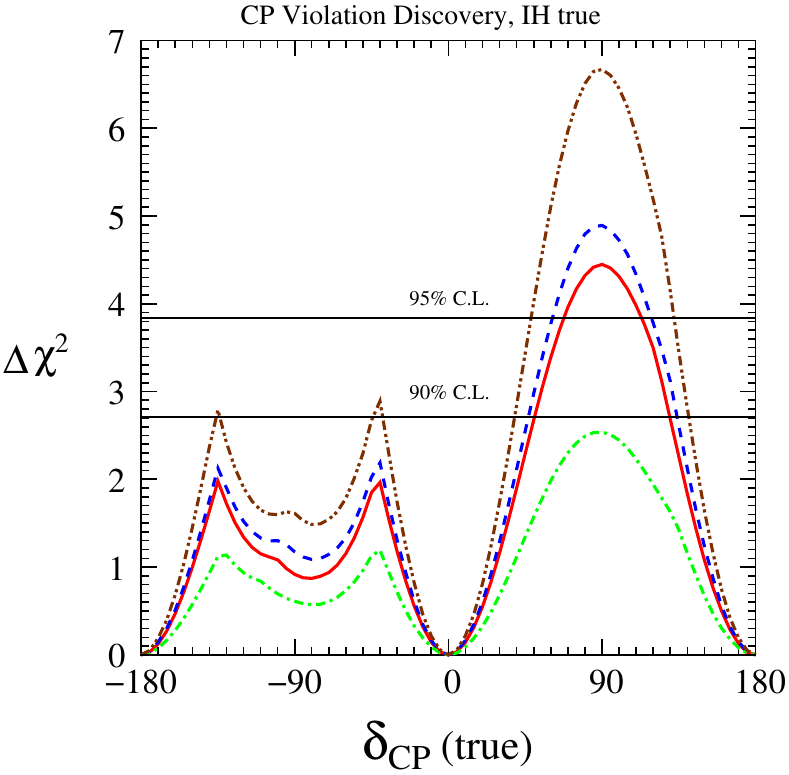}
\caption{{\footnotesize (colour online) CP violation discovery as a function of 
true value of $\dcp$ for the conservative values of input parameters, 
of new NO$\nu$A and of LArTPC of different masses. 
Left (right) panel is for NH (IH) as true hierarchy.}}
\label{fig:cpvdiscoverylarnova2}
\end{figure}

Figures~(\ref{fig:mhdiscovery-LAr-NOvA2}) 
and~(\ref{fig:cpvdiscoverylarnova2}) show, respectively,
the mass hierarchy and the CP violation sensitivities of
the new NO$\nu$A and the LArTPC of different masses, for
the conservative parameters (Best-fit 2). These figures 
are to compared with the corresponding 
figures~(\ref{fig:mhdiscovery-LAr-NOvA}) 
and~(\ref{fig:cpvdiscoverylarnova}). We find that both the
sensitivities are, in general, worse. 
 
\begin{figure}[H]
\centering
\includegraphics[width=0.49\textwidth]{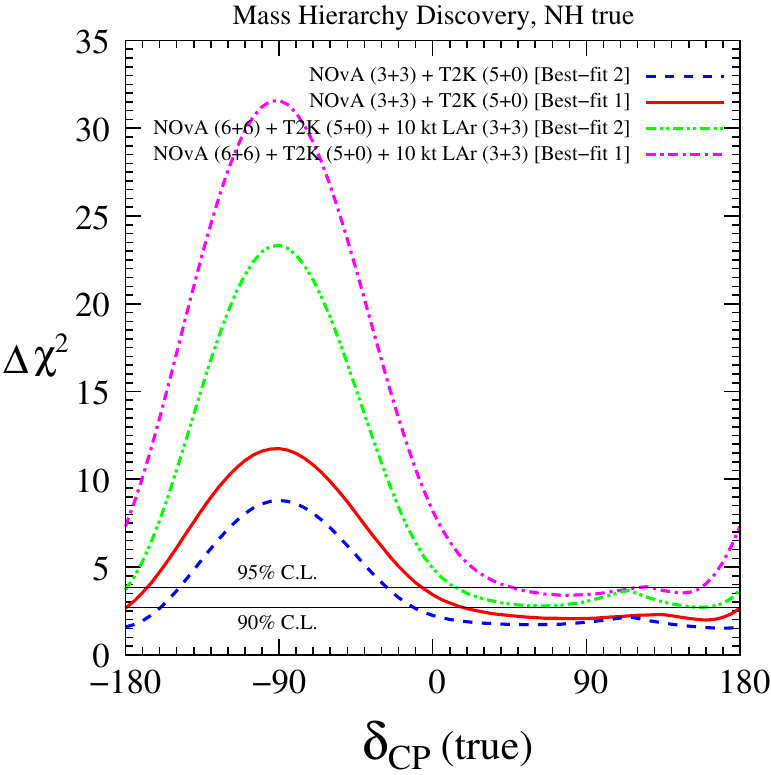}
\includegraphics[width=0.49\textwidth]{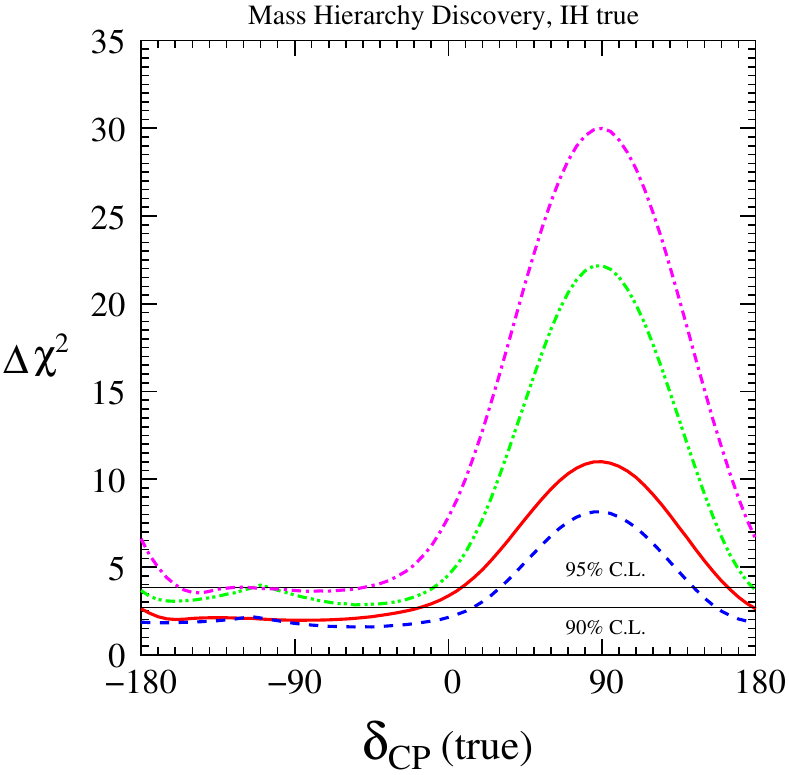}
\caption{{\footnotesize (colour online) Mass Hierarchy discovery as a function of 
true value of $\dcp$ for the conservative values of input parameters. 
Combined performance of new NO$\nu$A, 
T2K and LArTPC. Left (right) panel is for NH (IH) as true hierarchy.}}
\label{fig:mhdiscoverycomb2}
\end{figure}

Table~\ref{tab:compare2} lists the fraction of $\dcp$ values for which
we can obtain mass hierarchy sensitivity and CP violation sensitivity at
$90 \%$ C.L. and at $95 \%$ C.L. Comparing the corresponding entries
in Table~\ref{tab:compare} and~\ref{tab:compare2}, we find that the
mass hierarchy sensitivity is worse for all the combinations with one
exception. A $90\%$ C.L. hint of mass hierarchy can be obtained for
all the $\dcp$ values if we have $(6 \nu + 6 \bar{\nu})$ run of NO$\nu$A,
$(5\nu +0)$ run of T2K and $(3 \nu + 3 \bar{\nu})$ run of a 10 kt 
LArTPC. But, for all other cases, the mass hierarchy sensitivity is
worse because the leading term in $P(\nu_\mu \to \nu_e)$ oscillation 
probability is down by $30\%$, causing the number of events also 
to be down by a similar fraction. This is also reflected in 
figure~(\ref{fig:mhdiscoverycomb2}), which compares the combined
sensitivities for the two sets of parameters.

\begin{figure}[H]
\centering
\includegraphics[width=0.49\textwidth]{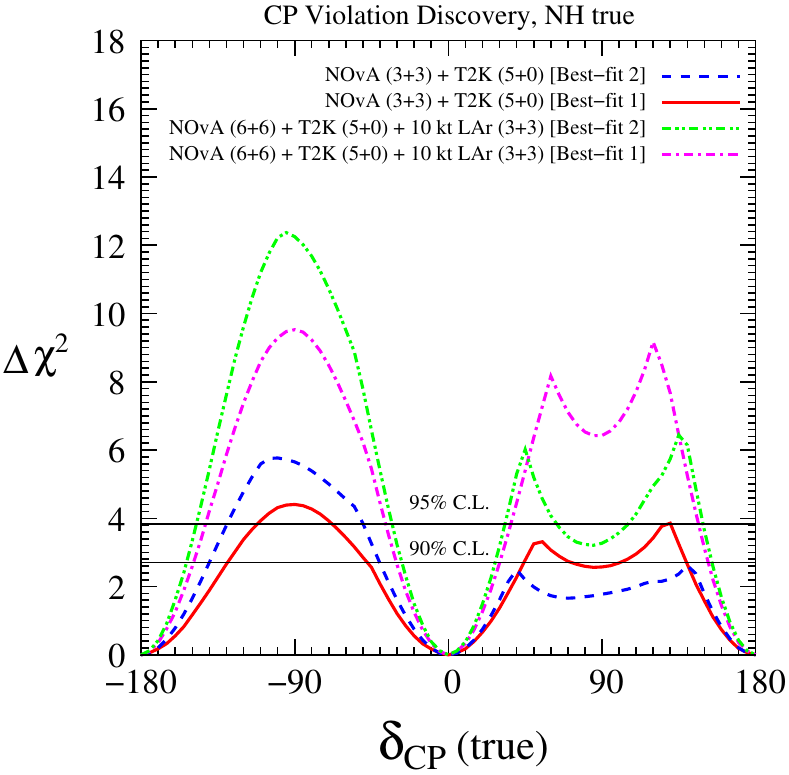}
\includegraphics[width=0.49\textwidth]{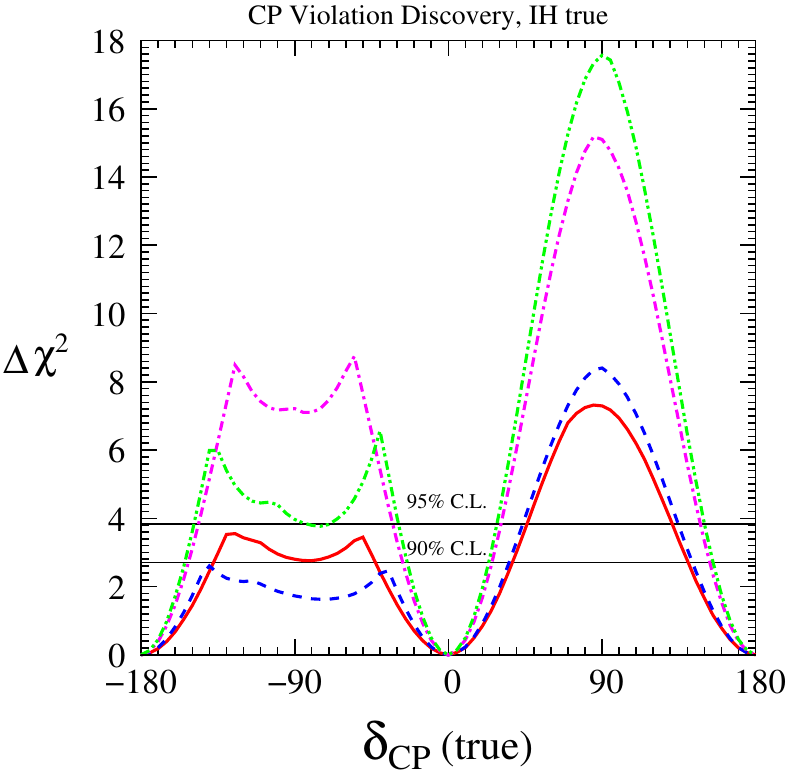}
\caption{{\footnotesize (colour online) CP violation discovery as a function of 
true value of $\dcp$ for the conservative values of input parameters. 
Combined performance of new NO$\nu$A, 
T2K and LArTPC. Left (right) panel is for NH (IH) as true hierarchy.}}
\label{fig:cpvdiscoverycomb2}
\end{figure}

Regarding CP violation sensitivity, there is a small improvement, 
with these conservative parameters. In this sensitivity, we are 
constrasting the situation of CP violation with $\dcp = 0$ and with 
$\dcp = 180^\circ$. Thus, in the numerator of $\Delta \chi^2$, the $30\%$ change 
in the leading term of $P(\nu_{\mu} \to \nu_e)$ cancels out. And the 
sub-leading term, which depends on $\dcp$, changes by less than $5\%$.
In the denominator of $\Delta \chi^2$, the leading term of $P(\nu_\mu \to
\nu_e)$ still dominates, thus the $\Delta \chi^2$ becomes larger. 
This is true for all the $\delta_{CP}$ values except for those in the range where
the marginalization over hierarchy brings $\Delta \chi^2$ down, 
as shown in figure~(\ref{fig:cpvdiscoverycomb2}).
Because of this, CP violation sensitivity becomes possible for a slightly
larger fraction of $\dcp$ values. 
In figure~(\ref{fig:cpvdiscoverycomb2}), comparison is done for only two 
of the setups but Table~\ref{tab:compare2} shows the sensitivities
with new inputs for all the setups.
The higher sensitivity to 
CP violation for smaller values of $\theta_{13}$ was noted before
in~\cite{Dick:1999ed, Donini:1999jc}.

\begin{table}[H]
\begin{center}
{\footnotesize
\begin{tabular}{|c|c|c||c|c|} \hline\hline
\multirow{4}{*}{Setups} & \multicolumn{4}{c|}{{\rule[0mm]{0mm}{6mm}Fraction of $\dcpt$}}
\cr\cline{2-5}
& \multicolumn{2}{c||}{{\rule[0mm]{0mm}{4mm}{\bf MH}}} & \multicolumn{2}{c|}{\rule[0mm]{0mm}{4mm}{{\bf CPV}}} 
\cr
\cline{2-5}
& NH true & IH true & NH true & IH true \cr
\hline
 NO$\nu$A (3+3) & 0.39 (0.33) & 0.37 (0.31) & 0.2 (0.1) & 0.22 (0.13) \cr
\hline
 NO$\nu$A (3+3) + T2K (5+0) & 0.41 (0.34) & 0.39 (0.31) & 0.28 (0.22) & 0.3 (0.25) \cr
\hline
 NO$\nu$A (6+6) + T2K (5+0) + 5 kt LArTPC (3+3) & 0.78 (0.5) & 0.89 (0.48) & 0.68 (0.45) & 0.71 (0.51) \cr
\hline
 NO$\nu$A (6+6) + T2K (5+0) + 10 kt LArTPC (3+3) & 1 (0.54) & 1 (0.54) & 0.7 (0.53) & 0.73 (0.63) \cr
\hline\hline
\end{tabular}
}
\caption{{\footnotesize Fractions of true values of $\dcp$ for which a discovery is possible for MH and CPV. 
The numbers without (with) parentheses correspond to 90\% (95\%) C.L. Here we take the central value for
$\sin^2 2 \theta_{13}$ to be 0.089 as predicted by Daya Bay. For $\sin^2 \theta_{23}$, the best fit value that we consider is 0.413. 
The results are shown for both NH and IH as true hierarchy.}}
\label{tab:compare2}
\end{center}
\end{table}

\end{appendix}

\bibliographystyle{JHEP}
\bibliography{neutosc1}

\end{document}